\documentclass[a4paper]{article}
\usepackage{a4,amsmath,cite}
\usepackage[dvips]{graphicx}
\usepackage{dcolumn}
\usepackage{caption}

\begin{document}

\large
\title{\textbf{Relativistic oscillator model with spin for nucleon resonances}}


\author{ D.A. Kulikov\thanks{kulikov\_d\_a@yahoo.com},
I.V. Uvarov\thanks{wvanj@pisem.net}, A.P. Yaroshenko
\\
\\
{\sl Theoretical Physics Department, FFEKS, Dniepropetrovsk
National
University }\\
{\sl 72 Gagarin avenue, Dniepropetrovsk 49010, Ukraine} }

\date{}
\maketitle

\abstract{The relativistic three-body problem is approached via the
extension of the $SL(2,C)$ group to the $Sp(4,C)$ one. In terms of
$Sp(4,C)$ spinors, a Dirac-like equation with three-body kinematics
is composed. After introducing the linear in coordinates
interaction, it describes the spin-1/2 oscillator. For this system,
the exact energy spectrum is derived and then applied to fit the 
Regge trajectories of baryon N-resonances in the $(E^2,J)$ plane.
The model predicts linear trajectories at high total energy $E$ with
some form of nonlinearity at low $E$. }

\normalsize

\section{Introduction}
\label{part1}

A relativistic equation for the symmetric quark model with harmonic
interaction was proposed by Feynman, Kislinger and Ravndal (FKR)
\cite{feynman} as far back as 1971. Since their work, the concept of
relativistic oscillator has been used for describing the spectra of
both the  ordinary hadrons \cite{mitra,toki,ishida,bohm} and
glueballs \cite{semay}. Of course, these models are purely
phenomenological ones. But, due to its exact solvability and
simplicity of the spectrum with levels grouped into shells, the
relativistic oscillator provides a convenient first approximation in
hadron systematics. It is especially important for light-quark
baryons where one should cope with a large amount of experimental
data on the excited states.

The original FKR model for baryons is based on the mass squared
operator  \cite{feynman}
\begin{equation}\label{eq0}
\mathcal{K}_\mathrm{FKR}=3(p_1^2+p_2^2+p_3^2)+\frac{1}{36}\Omega^2[(x_1-x_2)^2+
(x_2-x_3)^2+(x_3-x_1)^2]+C
\end{equation}
constructed from the four-momenta $p_i$ ($i=1,2,3$) of three
quarks and the conjugate positions $x_i$. Since the corresponding
eigenvalues are a succession of integers times $\Omega$, the mass
squared grows linearly with the angular momentum in general
agreement with experiment, but the price paid is the high
degeneracy of the spectrum. This degeneracy has been removed in
further algebraic approaches such as the interacting boson model
\cite{toki} and the stringlike collective model \cite{bijker},
which distinguish between excitations of different kinds.

It should be stressed that all the above models are formulated
assuming spinless quarks and thus they classify baryons according to
the orbital angular momentum $\ell$. However, in relativity, only
the total angular momentum $J$ and not its parts is defined. A
relativistic description of the light-baryon excitation spectrum in
terms of $J$ and parity was obtained in Ref.\cite{o4} by employing
the Lorentz group representations of the Rarita-Schwinger type.
Instead of the mass operator with an explicit interaction like
Eq.~(\ref{eq0}), there the Hamiltonian as a function of the Casimir
operators of the symmetry group is postulated. From the
representations involved, the authors of Ref.\cite{o4} infer that
their model has the spin content given by $J=\ell\pm 1/2$.

The question arises as to whether it is possible to incorporate
spin into the baryon model in an alternative and straightforward
way, by taking the "square root" from a second-order oscillator
operator. It is known that this is true in the case of the
one-body Klein-Gordon equation where one ends up with the Dirac
equation linear in both the momentum and the position, the
so-called Dirac oscillator model \cite{moshinsky}.

The purpose of this work is to construct a relativistic three-body
oscillator model based on an appropriate Dirac-like equation with
interaction. Our consideration makes use of the connection between
the Lorentz symmetry and the $SL(2,C)$ symmetry in the spinor
space. We apply the approach for composing relativistic wave
equations that starts with the extension of the $SL(2,C)$ group to
the symplectic $Sp(4,C)$ one \cite{kulikov}. In Ref.\cite{kulikov}
the fermion-boson problem was studied and an exactly solvable
model for the two-body oscillator with spin-1/2 was offered. In
the present work we generalize this model to the three-body case
and show that the exact solvability survives, but now the spectrum
has the $(J+N)$-degeneracy, which resembles the one seen in the
nucleon resonance spectrum.

The plan of the work is as follows. In Section 2 we perform the
symplectic space-time extension to obtain the relativistic
three-body kinematics. In Section 3 the three-body system with the
oscillator interaction involved through generalized momenta linear
in coordinates is studied. Section 4 is devoted to incorporating
the spin in the preceding results. Here we consider the Dirac-like
equation with the interaction and derive the corresponding energy
spectrum. This spectrum is applied in Section 5 to the description
of the nucleon excitations. Our conclusion is given in Section 6.

\section{Three-body kinematics based on the extension of the $SL(2,C)$
group} \label{part2}

In this Section we sketch out the procedure of the symplectic
space-time extension and apply it to construct a relativistic
operator with the three-body kinematics in spirit of the FKR
operator (\ref{eq0}).

Let us recall that the construction of relativistic wave equations
in the Minkowski space relies on the symmetry with the
$Sp(2,C)\equiv SL(2,C)$ group, which governs the transformations
of two-component Weyl spinors.  It is a universal covering group
for the homogeneous Lorentz group $SO(1,3)$. As a consequence,
there exists a one-to-one correspondence between $Sp(2,C)$
Hermitian spin-tensors of second rank and Minkowski four-vectors.
It allows one to parametrize the four-momentum of a relativistic
particle by the $Sp(2,C)$ Hermitian spin-tensor and to write down
the Dirac equation in terms of the Weyl spinors \cite{landau}.

In order to describe few-particle systems, we extend the
symplectic $Sp(2,C)$ group to the $Sp(4,C)$ one. This is the
minimal extension that preserves a non-degenerate antisymmetric
bilinear form $\eta_{\alpha\beta}=-\eta_{\beta\alpha}$
($\alpha,\beta=1,2,3,4$) in the spinor space.

Consider the $Sp(4,C)$ Hermitian spin-tensor
$\mathcal{P}_{\alpha\bar{\alpha}}$ (hereafter bared indices refer
to complex conjugate spinors). According to our previous analysis
\cite{kulikov}, it can be decomposed into four Minkowski
four-momenta as
\begin{equation}\label{eq8}
\mathcal{P}=I\otimes\sigma^m w_m+\tau^1\otimes\sigma^m p_m+
\tau^2\otimes\sigma^m u_m+\tau^3\otimes\sigma^m q_m
\end{equation}
where $w_m$, $p_m$, $u_m$, $q_m$ $(m= 0,1,2,3)$ are the Minkowski
four-momenta, and the following representation with $2\times 2$
unit matrix $I$ and the Pauli matrices $\tau^i$ is used
\begin{eqnarray}\label{eq4}
\sigma^0=I, \quad \sigma^1=\tau^1, \quad \sigma^2=\tau^2, \quad
\sigma^3=\tau^3.
\end{eqnarray}
Note that the second factor in the direct matrix products in
Eq.~(\ref{eq8}) is the $Sp(2,C)$ momentum spin-tensor, while the
first one is due to the group extension.

It should be stressed that the description of a three-body system
requires one time-like and nine space-like variables, whereas the
$Sp(4,C)$ momentum spin-tensor has sixteen components. However, we
are able to decrease the number of the independent components by
introducing subsiduary conditions in a $Sp(4,C)$-invariant form.

For deriving such conditions, we multiply the $Sp(4,C)$ momentum
spin-tensor $\mathcal{P}_{\alpha\bar{\alpha}}$ by the transposed
one $\tilde{\mathcal{P}}^{\bar{\alpha}\beta}$, to obtain the
Klein-Gordon-like operator
\begin{equation}\label{eq8a}
\mathcal{K}\equiv\mathcal{P}\tilde{\mathcal{P}}=w^2+p^2-u^2+q^2+\sum^{5}_{A=1}\gamma_A
K^A
\end{equation}
where $w^2=(w^0)^2-\mathbf{w}^2$, $p^2=(p^0)^2-\mathbf{p}^2$ etc,
$\gamma_A$ are direct products of the Pauli matrices, and $K^A$
are quadratic forms with respect to the four-momenta.

Five quantities $K^A$ are components of a complex vector that
transforms according to the representation $SO(5,C)\subset
Sp(4,C)$. To restore the diagonal form of the Klein-Gordon
operator, we put $K^A=0$ on wave functions. Such a condition is
invariant under the $Sp(4,C)$ group transformations because if a
vector equals to zero in one frame, then it equals to zero in all
frames.

Being written in terms of the four-momenta, the imposed equality
$K^A=0$ reads
\begin{eqnarray}\label{eq8b}
&&wp+pw=0,\qquad wq+qw=0,\qquad up+pu=0,\qquad uq+qu=0, \nonumber \\
&&u^m w^n + w^n u^m - u^n w^m - w^m u^n-\epsilon^{mnkl}(p_k q_l +
q_l p_k)=0
\end{eqnarray}
where $\epsilon^{mnkl}$ is the totally antisymmetric tensor
($\epsilon^{0123}=+1$).

These conditions imply that either one or three of the
four-momenta $w_m$, $p_m$, $u_m$ and $q_m$ must transform as axial
vectors. Let $u_m$ be the sole axial vector. Then we may connect
the remaining four-momenta with the four-momenta $p_{1}^m$,
$p_{2}^m$ and $p_{3}^m$ of the constituent particles in the
standard manner \cite{feynman}
\begin{equation}\label{eq9}
w^m=p_{1}^m+p_{2}^m+p_{3}^m ,\quad
p^m=\sqrt{\frac{3}{2}}\;(p_{1}^m-p_{2}^m), \quad
q^m=\frac{1}{\sqrt{2}}\;(p_{1}^m+p_{2}^m-2p_{3}^m).
\end{equation}

Supposing the total four-momentum $w_m$ to be conserved, for an
arbitrary four-vector $a_m$ we can introduce its transverse and
longitudinal, with respect to $w_m$, parts
\begin{equation}\label{eq185}
a_{\bot}^m=(g^{mn}-w^m w^n/w^2)a_n, \qquad a_{\parallel}^m=(w^m
w^n/w^2)a_n
\end{equation}
where $g^{mn}=diag(1,-1,-1,-1)$ is the Minkowski metrics.

With this notation, the subsiduary conditions (\ref{eq8b}) are
reduced to
\begin{equation}\label{eq19}
p^{m}_{\parallel}=0, \qquad q^{m}_{\parallel}=0, \qquad u_{\bot
m}=\frac{1}{w^2}\epsilon_{mnkl}w^n p^k q^l.
\end{equation}
From the first two equalities it becomes evident that the relative
time variables are  removed, as it is necessary for the three-body
problem \cite{droz06}. The last equality shows that the axial
$u_m$ is an auxiliary quantity and only its longitudinal  part
$u^{m}_{\parallel}$ remains independent.

Upon inserting Eqs.~(\ref{eq9}) and (\ref{eq19}), the
Klein-Gordon-like operator (\ref{eq8a}) takes the form
\begin{equation}\label{eq10}
\mathcal{K}=3(p_1^2+p_2^2+p_3^2)-u^{2}_{\parallel}
+\frac{1}{w^2}[p^{2}_{\bot}q^{2}_{\bot}-(p_{\bot}q_{\bot})^2],
\end{equation}
to be compared with the kinetic part of the FKR operator
(\ref{eq0}). We see that the yet undetermined quantity
$u^{2}_{\parallel}$ substitutes the additive constant $C$
introduced in the FKR model to account for such effects as the
difference of quark masses. In our model, intended to describe the
nucleon resonances only, we put $u^{m}_{\parallel}=0$. The last
term in Eq.~(\ref{eq10}) has no analog in the FKR operator. But
this term vanishes in the non-relativistic limit where $w^2$
containing the rest energy dominates over space-like
$p^{m}_{\bot}$ and $q^{m}_{\bot}$.

Thus, within the approach based on the extension of the $SL(2,C)$
group, the kinematics of three non-interacting particles is
described by the Klein-Gordon-like operator (\ref{eq10})
supplemented with the subsiduary conditions (\ref{eq19}). Mention
that the two-body kinematics can be obtained now by imposing the
further restriction $q^{m}=0$ (see Ref.\cite{kulikov} for
details).

\section{Oscillator interaction}
\label{part3}

Now we are going to include the interaction in the description.
This can be made by replacing the four-momenta of particles by the
generalized momenta ($p_i^m\rightarrow\pi_i^m=p_i^m-A_i^m$,
$i=1,2,3$), so that each particle is in an external potential of
the others. We assume that both the Klein-Gordon-like operator
(\ref{eq10}) (properly symmetrized) and the subsiduary conditions
(\ref{eq19}) are subject to these replacements.

Because the generalized momenta do not, generally, commute with
each other, the question on the compatibility arises. In the
language of the Dirac's quantum mechanics with constraints,
Eqs.~(\ref{eq19}) and (\ref{eq10}) are the first-class
constraints. Then a sufficient condition of the compatibility
implies that their mutual commutators do vanish without producing
second-class constraints.

We choose the simplest generalized momenta that meet the above
compatibility requirement and satisfy the subsiduary conditions of
the same form (\ref{eq8b}) as in the non-interacting case. Namely,
this is the interaction linear in the coordinates of particles in
spirit of the Dirac oscillator model \cite{moshinsky}
\begin{eqnarray}\label{eq20}
&&\pi^m_1=p^m_1-\frac{\lambda}{3\sqrt{3}}
(x^m_{2\bot}-x^m_{3\bot}), \quad
\pi^m_2=p^m_2-\frac{\lambda}{3\sqrt{3}} (x^m_{3\bot}-x^m_{1\bot}),
\nonumber \\
&&\phantom{\pi^m_1=p^m_1-} \pi^m_3=p^m_3-\frac{\lambda}{3\sqrt{3}}
(x^m_{1\bot}-x^m_{2\bot})
\end{eqnarray}
where $\lambda$ is a coupling constant.

In terms of the relative momenta Eq.~(\ref{eq20})  translates into
\begin{equation}\label{eq20a}
p^m\rightarrow P^m=p^m-\lambda y_{\bot}^m, \qquad q^m\rightarrow
Q^m=q^m+\lambda x_{\bot}^m
\end{equation}
with the relative positions defined as the Jacobi coordinates
\begin{equation}\label{eq21}
x^m=\frac{x_{1}^m-x_{2}^m}{\sqrt{6}}, \qquad
y^m=\frac{x_{1}^m+x_{2}^m-2x_{3}^m}{3\sqrt{2}},
\end{equation}
which obey $[p^m,x^n]=\mathrm{i}g^{mn}$,
$[q^m,y^n]=\mathrm{i}g^{mn}$.

As a consequence, the following commutation relation holds
 \begin{equation}\label{eq21a}
[P^m,Q^n]=2\mathrm{i}\lambda (g^{mn}- w^m w^n/w^2)
\end{equation}
that resembles the commutator of the generalized momenta for a
charged particle in the magnetic field. Associated with the Landau
levels, the latter system is indeed tightly connected with the
harmonic oscillator.

To separate oscillator degrees of freedom in our three-body
system, let us consider its three-dimensional reduction. Although
it may be performed in a covariant manner, by using the
decomposition (\ref{eq185}), we prefer a more illustrative
approach and pass to the center-of-mass (CM) frame in which
$\mathbf{w}=0$. Then $E=w^0$ is the total energy and the dynamics
of the relative motion is described by the three-dimensional
coordinates $\mathbf{x}_{\bot}=\mathbf{x}$ and
$\mathbf{y}_{\bot}=\mathbf{y}$.

From Eq.~(\ref{eq19}) it follows that $P^0=Q^0=0$ and the
Klein-Gordon-like operator (\ref{eq10}), rewritten through the
vectors of generalized relative momenta $\mathbf{P}$ and
$\mathbf{Q}$, becomes
\begin{equation}\label{eq210}
-\mathcal{K}=\mathbf{P}^2+\mathbf{Q}^2-E^2+\frac{1}{E^2}(\mathbf{Q}\times\mathbf{P})^2.
\end{equation}
If one now introduces the creation and annihilation operators
\begin{eqnarray}\label{eq230}
\mathbf{c}=\frac{\mathbf{Q}+\mathrm{i}\mathbf{P}}{2\sqrt{\lambda}},
\qquad
\mathbf{c}^{\dag}=\frac{\mathbf{Q}-\mathrm{i}\mathbf{P}}{2\sqrt{\lambda}},
\end{eqnarray}
the first two terms in the right-hand side of Eq.~(\ref{eq210})
can be identified with the operator counting oscillator quanta,
namely,
\begin{equation}\label{eq240}
\mathbf{P}^2+\mathbf{Q}^2=4\lambda\left(\mathbf{c}^{\dag}\cdot\mathbf{c}
+\frac{3}{2}\right).
\end{equation}
As for the last term in Eq.~(\ref{eq210}), it commutes with
$\mathbf{P}^2+\mathbf{Q}^2$ and thus represents a relativistic
correction that does not change the number of quanta.

It should be emphasized that the model under consideration
exploits its degrees of freedom in a different way than the FKR
model. Within the latter, all six degrees of freedom associated
with the relative motion are of the harmonic oscillator nature. In
contrast to this, our model obviously possesses only one
three-dimensional oscillator mode through the operators
$\mathbf{c}$ and $\mathbf{c}^{\dag}$.

To reveal the three other degrees of freedom, we look at the orbital
angular momentum operator for the system $\mathbf{l}$. In the
presence of the interaction, its elements can be rearranged as
\begin{equation}\label{eq242}
\mathbf{l}\equiv\mathbf{x}\times\mathbf{p}+\mathbf{y}\times\mathbf{q}=\mathbf{M}+\mathbf{N}
\end{equation}
where
\begin{equation}\label{eq244}
\mathbf{M} =
\frac{1}{2\lambda}(\mathbf{q}+\lambda\mathbf{x})\times
(\mathbf{p}-\lambda\mathbf{y})\equiv\frac{1}{2\lambda}\mathbf{Q}\times\mathbf{P},
\qquad \mathbf{N}
=-\frac{1}{2\lambda}(\mathbf{q}-\lambda\mathbf{x})\times
(\mathbf{p}+\lambda\mathbf{y})
\end{equation}
both obey the algebra for angular momenta, $[L_a ,
L_b]=\mathrm{i}\epsilon_{abc}L_c$ $(a,b,c=1,2,3)$, $\mathbf{L}
=\mathbf{M}$ or $\mathbf{N}$. But $\mathbf{M}$ also satisfies the
commutation relations with $\mathbf{P}$ and $\mathbf{Q}$ thought
as a linear momentum and a position respectively
\begin{equation}\label{eq246}
[M_a , P_b]=\mathrm{i}\varepsilon_{abc}P_c, \qquad [M_a ,
Q_b]=\mathrm{i}\varepsilon_{abc}Q_c.
\end{equation}

As for $\mathbf{N}$, denoting its parts by
\begin{equation}\label{eq20b}
\mathbf{Q}^{(-)}=\mathbf{q}-\lambda\mathbf{x},  \qquad
\mathbf{P}^{(-)}=\mathbf{p}+\lambda\mathbf{y},
\end{equation}
we get the commutation relations
\begin{equation}\label{eq246a}
[N_a , P^{(-)}_b]=\mathrm{i}\varepsilon_{abc}P^{(-)}_c, \qquad [N_a
, Q^{(-)}_b]=\mathrm{i}\varepsilon_{abc}Q^{(-)}_c,
\end{equation}
which are identical to (\ref{eq246}) but contain $\mathbf{N}$,
$\mathbf{P}^{(-)}$ and $\mathbf{Q}^{(-)}$ that commute with all the
quantities entering the oscillator Klein-Gordon-like operator
(\ref{eq210}) and, hence, are conserved. This commutativity property
suggests that the oscillator model under consideration does have the
special Euclidian $SE(3)$ symmetry generated by $\mathbf{N}$ and,
say, $\mathbf{P}^{(-)}$, in addition to the $U(2)$ symmetry
generated by the number of quanta operator
$\mathbf{c}^{\dag}\cdot\mathbf{c}$ and the angular momentum
$\mathbf{M}=\mathrm{i}\mathbf{c}\times\mathbf{c}^{\dag}$. The latter
$U(2)$ is obviously a remnant of the $U(3)$ symmetry of the ordinary
harmonic oscillator \cite{elliott}, which has been partially broken
by the last term in our oscillator Klein-Gordon-like operator
(\ref{eq210}).

In its turn, the $SE(3)$ algebra is known as the symmetry of the
free particle motion and its irreducible representations labeled by
the linear and angular momenta squared correspond to the spherical
waves. In the present model the three non-oscillator degrees of
freedom associated with such spherical waves contribute only through
the orbital angular momentum operator
$\mathbf{l}=\mathbf{M}+\mathbf{N}$ where $\mathbf{N}$ belongs to the
$SE(3)$ algebra. An explicit effect is that, calculating the
eigenvalues of $\mathbf{M}^2$, one obtains the sequence of the
values $M=\ell+N,\ell+N-1,...,|\ell-N|$. Therefore the energy
spectrum of the system, to be derived from Eq.~(\ref{eq210}) that
involves $\mathbf{M}$ but not $\mathbf{l}$ and $\mathbf{N}$, will
contain a plenty of states with the same orbital number $\ell$ and
the number of oscillator quanta. We can overcome this shortcoming by
appealing to the analogy with the flux-tube model of hadrons (see
Ref.\cite{Selem} for contemporary discussion), in which the
flux-tube rotation results in the linear growth of energy squared
versus orbital angular momentum (the famous Chew-Frautschi
conjecture). We thus postulate that the non-oscillator degrees of
freedom may only increase the energy, by picking up the maximal
value of the angular momentum $M=\ell+N$ from the sequence.

Keeping in mind the application to the baryon spectrum, we treat
the above partition of the degrees of freedom as due to the
emergence of a diquark, two-quark cluster, in the interacting
three-quark system. It is known that existence of diquarks is
supported by various models (see Ref.\cite{diquark} for review)
and the quark-diquark picture provides a simpler classification of
the excited nucleon states \cite{anisovichb}. From this point of
view, the three oscillator degrees of freedom of our system in the
CM frame correspond to the interaction between a quark and a
diquark, while the three remaining ones to rotational excitations.

\section{Oscillator with spin and its energy spectrum}
\label{part4}

In order to describe baryons, we shall incorporate spin in the
preceding results. We take the "square root"\; from the
Klein-Gordon-like operator to obtain the Dirac-like equation. Then
the analytical formulae for eigenenergies of the three-body system
with spin are derived.

\subsection{Dirac-like three-body equation}

Consider the system with spin equal to $1/2$. In practice, this will
be a baryon consisting of a spin-1/2 quark and a spinless (or "good"
in the standard terminology) diquark. The wave function of this
system can be represented by a Dirac bispinor or a pair of Weyl
spinors. Using the four-component $Sp(4,C)$ Weyl spinors
$\varphi_{\alpha}$, $\bar{\chi}^{\bar{\alpha}}$ and the momentum
spin-tensor $\mathcal{P}_{\alpha\bar{\alpha}}$ given by
Eq.~(\ref{eq8}), one may compose the wave equation
\begin{equation}\label{eq7}
\mathcal{P}_{\alpha\bar{\alpha}}\bar{\chi}^{\bar{\alpha}}=m\varphi_{\alpha},
\qquad
\tilde{\mathcal{P}}^{\bar{\alpha}\alpha}\varphi_{\alpha}=m\bar{\chi}^{\bar{\alpha}},
\end{equation}
with  $m$ being a mass parameter.

Assuming the same oscillator interaction (\ref{eq20a}) as in the
previous Section, in the CM frame Eq.~(\ref{eq7}) reduces to
\begin{eqnarray}\label{eq250}
&&\left(E-\tau^1\otimes\boldsymbol\tau\cdot\mathbf{P}
-\tau^3\otimes\boldsymbol\tau\cdot\mathbf{Q}-\frac{2\lambda}{E}\tau^2\otimes
\boldsymbol\tau\cdot\mathbf{M}\right)\bar{\chi}=m\varphi, \nonumber \\
&&\left(E+\tau^1\otimes\boldsymbol\tau\cdot\mathbf{P}
+\tau^3\otimes\boldsymbol\tau\cdot\mathbf{Q}-\frac{2\lambda}{E}\tau^2\otimes
\boldsymbol\tau\cdot\mathbf{M}\right)\varphi=m\bar{\chi},
\end{eqnarray}
which can be brought conveniently to a Hamiltonian form. By
introducing
\begin{equation}\label{eq252}
\Psi=\left(\begin{array}{c}
       \psi_{+} \\
       \psi_{-} \\
     \end{array}\right)=
     \left(\begin{array}{c}
       (\bar{\chi}+\varphi)/\sqrt{2} \\
       (\bar{\chi}-\varphi)/\sqrt{2} \\
     \end{array}\right),
\end{equation}
we get the equation with the energy-dependent Dirac-like
Hamiltonian
\begin{equation}\label{eq254}
H\Psi=E\Psi, \,\, H=(\tau^1\otimes\boldsymbol\tau\cdot\mathbf{P}
+\tau^3\otimes\boldsymbol\tau\cdot\mathbf{Q})\left(\begin{array}{cc}
       0 & 1\\
       1 & 0\\
     \end{array}\right)+m\left(\begin{array}{cr}
       1 & 0\\
       0 & -1\\
     \end{array}\right)+\frac{2\lambda}{E}\,\tau^2\otimes\boldsymbol\tau\cdot\mathbf{M}.
\end{equation}

Now it is crucial to find a complete set of operators that commute
with the Hamiltonian and among themselves. Their eigenvalues will
supply us with quantum numbers labeling a state of the system.

It is easy to verify that the total angular momentum
\begin{equation}\label{eq256}
\mathbf{J}=\mathbf{x}\times\mathbf{p}+\mathbf{y}\times\mathbf{q}+I\otimes\frac{\boldsymbol{\tau}}{2}\equiv
\mathbf{M}+\mathbf{N}+I\otimes\frac{\boldsymbol{\tau}}{2}
\end{equation}
commutes with $H$. Moreover, the operator $\mathbf{N}^2$ commutes
with both $H$ and $\mathbf{J}^2$. This amounts to say that, apart
from the total spin $J$ defined by $\mathbf{J}^2\Psi=J(J+1)\Psi$,
there must exist another good quantum number $N$ associated with
rotational degrees of freedom through $\mathbf{N}^2\Psi=N(N+1)\Psi$.

Like the above-discussed spinless case, the wave equation involves
only a part of the total angular momentum -- it is
$\mathbf{j}=\mathbf{M}+I\otimes\boldsymbol{\tau}/2$ now. Invoking
the same analogy with the flux-tube model as in the end of the
previous section, we pick up the maximal eigenvalue of
$\mathbf{j}^2$ for the eigenenergies to depend on. That is, we
select the states with $j=J+N$. It should be added that $N$ may
attain only integer values $0,1,...$ and not half-integer ones
because $\mathbf{N}$ is the differential operator in the coordinate
space.

The set of mutually commuting operators also includes the angular
momenta projections $J_3$ and $N_3$, the spherical wave number
squared $\mathbf{P}^{(-)2}$ that does not affect the energy
spectrum, the spin-orbit coupling operator $\kappa$ and the constant
matrix $\varsigma$ given by
\begin{equation}\label{eq258}
\kappa=I\otimes(\boldsymbol\tau\cdot\mathbf{M}+I)\left(\begin{array}{cr}
1 & 0\\
      0 & -1\\
     \end{array}\right),\qquad \varsigma=\tau^2\otimes I \left(\begin{array}{cr}
1 & 0\\
      0 & -1\\
     \end{array}\right).
\end{equation}

The matrix $\varsigma$ reflects the doubling of the number of
spinor components versus the ordinary Dirac equation. It can be
checked that, by applying the projectors $(1\pm \varsigma)/2$, the
three-body equation (\ref{eq254}) is split into two separate ones
for two four-component Dirac bispinors. In view of this doubling
it is temptimg to interpret the symmetry with respect to the
unitary transformation generated by $\varsigma$ as a remnant of
the isospin symmetry between the proton and the neutron.
Remarkably, the symmetry is not broken by the oscillator
interaction.

Now let us derive an explicit oscillator-type equation for our
system. This can be made by taking the square of the Hamiltonian
$H$ and then subtracting an appropriate term linear in $H$, to
diagonalize the matrix. Namely, we evaluate $[H^2-(4/E)\lambda
\varsigma \kappa H]\Psi$, to get the second-order equation
\begin{equation}\label{eq259}
(\mathbf{P}^2+\mathbf{Q}^2)\psi_{\pm}=\left[E^2-\left( m-
\frac{2\lambda\varsigma}{E}\right)^2 \pm 2\lambda\varsigma
+\frac{4\lambda^2 \kappa^2}{E^2} \right]\psi_{\pm}
\end{equation}
that describes a harmonic oscillator with the additional spin-orbit
interaction, which enters via the operator $\kappa$.

\subsection{Energy spectrum}

We are in the position to calculate the energy spectrum of the
system. First, it should be noticed that in the last equation the
operator $\varsigma$, whose eigenvalues are $\pm 1$, is
accompanied with the coupling constant $\lambda$. We therefore
absorb $\varsigma$ into the definition of $\lambda$ and look for
two branches of the spectrum corresponding to $\lambda>0$ and
$\lambda<0$, respectively.

The problem merely reduces to expressing the eigenvalues of the
operators involved in Eq.~(\ref{eq259}) in terms of the observed
spin-parity $J^P$. The left-hand side is just the operator
(\ref{eq240}) counting the number of the oscillator quanta. In view
of the algebra (\ref{eq246}), this number can be partitioned in the
standard manner for the isotropic oscillator \cite{elliott} as
$(2n+M)$ where $n=0,1,...$ is the radial quantum number and
$M=0,1,...$ is the orbital quantum number defined by
$\mathbf{M}^2\psi=M(M+1)\psi$, with $\psi$ standing for one of
$\psi_{+}$ and $\psi_{-}$ from the decomposition (\ref{eq252}).

Since $\mathbf{M}^2$ is not conserved, $\psi_{+}$ and $\psi_{-}$
refer to different values of $M$. From Eqs.~(\ref{eq256}) and
(\ref{eq258}) we deduce that these values and also the parity are
unambiguosly determined by the conserved total angular momentum
$J$ and the eigenvalue $\kappa=\pm (J+N+1/2)$ of the spin-orbit
coupling operator as
\begin{equation}\label{eq260}
M=J+N\mp\frac{\kappa}{2|\kappa|}, \qquad
P=(-1)^{J+N-\kappa/2|\kappa|}
\end{equation}
where the upper (lower) sign inside $M$ refers to $\psi_{+}$
($\psi_{-}$).

Collecting (\ref{eq240}), (\ref{eq259}) and (\ref{eq260}) we
arrive at the dispersion relations
\begin{eqnarray}\label{eq270}
\left(E-\frac{2|\lambda|(J+N+1/2)}{E}\right)^2-\left( m-
\frac{2\lambda}{E}\right)^2=4|\lambda|\left(2n+\frac{|\lambda|-\lambda}{2|\lambda|}\right),
\; P=(-1)^{J+N-1/2}, 
\nonumber \\
\left(E-\frac{2|\lambda|(J+N+1/2)}{E}\right)^2-\left( m-
\frac{2\lambda}{E}\right)^2=4|\lambda|\left(2n+\frac{|\lambda|+\lambda}{2|\lambda|}\right),
\; P=(-1)^{J+N+1/2}. \nonumber \\ 
\end{eqnarray}
Here the first and second lines were obtained by using the
equations for $\psi_{+}$ and $\psi_{-}$, respectively.

The ground-state ($n=0$) solutions for which one of the components
vanishes need a special care. As seen from the structure of the
Hamiltonian (\ref{eq254}), such solutions are obtained by setting
$(\tau^1\otimes\boldsymbol\tau\cdot\mathbf{P}
+\tau^3\otimes\boldsymbol\tau\cdot\mathbf{Q})\psi=0$ and, if
present, must have energy in agreement with the general formulae
(\ref{eq270}).

In the case of $\lambda>0$, the last equation admits a normalizable
solution for $\psi_{+}$ and the corresponding ground state with
$\psi_{-}=0$ is characterized by
\begin{equation}\label{eq280}
\left(E-\frac{2\lambda(J+N+1/2)}{E}\right)-\left( m-
\frac{2\lambda}{E}\right)=0, \quad P=(-1)^{J+N-1/2} \quad
(\lambda>0)
\end{equation}
that agrees with the first line of Eqs.~(\ref{eq270}) with $n=0$.

In the case of $\lambda<0$, there exists the ground state with
$\psi_{+}=0$ possessing
\begin{equation}\label{eq290}
\left(E-\frac{2|\lambda|(J+N+1/2)}{E}\right)+\left( m-
\frac{2\lambda}{E}\right)=0, \quad P=(-1)^{J+N+1/2} \quad
(\lambda<0)
\end{equation}
that falls into the second line of Eqs.~(\ref{eq270}).

Before writing down explicit solutions to the derived equations, let
us notice that the cases of  $\lambda>0$ and $\lambda<0$ transform
one into each other under the simultaneous change
$\lambda\rightarrow -\lambda$, $m\rightarrow -m$ and $P\rightarrow
-P$. The inversion of $P$ does not matter because only the relative
parity can be defined for fermions. Thus the situation resembles a
classical model of symmetry breakdown: a point-like classical
particle moving on the line, under the sole influence of the
w-shaped potential $V (x) = (x-a)^2 (x+a)^2$. In this model there
are two positions of stable equilibrium, $x=a$ and $x=-a$, which are
transmuted one into each other by the parameter redefinition,
$a\rightarrow -a$. However, only one of them has to be selected to
get a physical picture. Coming back to the three-body model, we
suppose its spectrum is in the Nambu-Goldstone mode, so that only
one of the branches with $\lambda>0$ and $\lambda<0$ is
spontaneously selected. For definiteness, we put $\lambda>0$ in what
follows.

The solution to the ground state ($n=0$)  equation (\ref{eq280}) is given by
\begin{equation}\label{eq295}
 J=\frac{E^2-mE}{2\lambda}-N+\frac{1}{2}, \quad P=(-1)^{J+N-1/2}
\end{equation}
that can be viewed as a Regge trajectory, \textit{i.e.} the
Chew-Frautschi plot of the total angular momentum $J$ versus total
energy squared $E^2$ over a set of particles whose other quantum
numbers are fixed (beyond the strict S-matrix concept of Regge
poles). This Regge trajectory is nearly linear in $E^2$ at high $E$,
as usually expected in the hadron spectroscopy. It is easy to verify
that being treated as the square equation with respect to $E$
Eq.~(\ref{eq295}) has a single positive root and thus no ambiguity
in extracting energy from it may appear.

The solutions to Eqs.~(\ref{eq270}) written in the form of the Regge
trajectories for the states with $n=1,2,...$ read as
\begin{equation}\label{eq301}
{\displaystyle
  J=\frac{E^2}{2\lambda}-N-\frac{1}{2}-\sqrt{\frac{2nE^2}{\lambda}
+\left(\frac{mE}{2\lambda}-1\right)^2}, } \qquad P=(-1)^{J+N-1/2},
\end{equation}
\begin{equation}\label{eq302}
{\displaystyle
 J=\frac{E^2}{2\lambda}-N-\frac{1}{2}-\sqrt{\frac{(2n+1)E^2}{\lambda}
+\left(\frac{mE}{2\lambda}-1\right)^2},} \qquad P=(-1)^{J+N+1/2}.
\end{equation}
Here the negative sign in front of the square root was chosen to
assure the growth of $E$ with the increase in $n$. Indeed, the
growth of $E$ corresponds to pulling the graph $J(E^2)$ downward in
the $(E^2,J)$ plane.

On the other hand, one may treat each line of Eqs.~(\ref{eq270}) as
an equation of the order four in $E$ and apply the Descartes'{} rule
of signs. This rule states that the number of positive real roots of
an algebraic equation with real coefficients
$a_{k}x^{k}+\cdots+a_{1}x +a_{0}=0$ is never greater than the number
of changes of signs in the sequence $a_{k},\ldots,a_{1},a_{0}$ (not
counting the null coefficients) and, if less, then always by an even
number. Using this rule, one can prove that the above-mentioned
equation for energies has no more than two positive roots. However,
as seen from Eqs.~(\ref{eq301}),(\ref{eq302}), the corresponding
Regge trajectory $J(E^2)$ is a monotonously increasing function for
high enough $E$. Hence, one can extract energy from it
unambiguously.

It is worth adding that the obtained spectrum is similar to that of
the two-body oscillator with spin which we considered in
Ref.~\cite{kulikov} (see Eqs.~(20) there, note that $m$ and
$\lambda$ were rescaled). The main difference is the presence of the
new quantum number $N$ that has no analog in the two-body case.
Thus, if the two-body oscillator of Ref.~\cite{kulikov} may be
viewed as a quark-diquark model with a rigid diquark, the present
three-body treatment accounts for some extra excitations deforming
the diquark. In the next Section we check the applicability of the
three-body model, by applying it to describe the spectrum of nucleon
resonances.

\section{Application. Regge trajectories of nucleon resonances}
\label{part5}

We shall describe the $N$-resonance states and omit the
$\Delta$-resonances. The reason is that in the quark-diquark picture
the ground state $\Delta(1232)$ as well as its radial excitations
correspond to the spin $S=3/2$ and thus can hardly be accommodated
by the Dirac bispinor transforming according to the
$(1/2,0)\oplus(0,1/2)$ representation of the group
$SL(2,C)=Sp(2,C)\subset Sp(4,C)$ we use.

To start with, consider the nucleon $N(940)$, the lightest state
having $J^P={1/2}^{+}$. From  Eq.~(\ref{eq295}) it follows that the
energy of the ground state with $N=0$ and $J^P={1/2}^{+}$ is $E=m$,
{\it i.e.}, the value of the mass parameter $m=0.940\,$GeV is
unambiguously determined by the nucleon mass. The remaining slope
parameter $\lambda=0.345\,$GeV$^2$ was chosen so as to fit the
nucleon Regge trajectory that also contains the well-established
states $N_{5/2+}(1680)$ and $N_{9/2+}(2220)$. We did not distinguish
between this trajectory and that for the negative-parity states
$N_{3/2-}(1520)$ and $N_{7/2-}(2190)$. Likewise, all other
trajectories were thought to contain the states with $J=1/2$, $3/2$,
$5/2$,... and the alternating parity.

It should be pointed out that within our model the Regge
trajectories can be obtained in three different ways. First, one can
get the opposite-parity states with $J^P=1/2^-$, $3/2^+$,
$5/2^-$,..., by switching from Eq.~(\ref{eq301}) with
$P=(-1)^{J+N-1/2}$ to Eq.~(\ref{eq302}) with $P=(-1)^{J+N+1/2}$.
Next, there exist radial excitations with $n=1,2,...$ Last, one
should consider the Regge trajectories with $N=1,2,...$ which are
obtained by shifting the $N=0$ trajectories downward to 1,2,...
units in $J$. It is the trajectories of this third type, generated
from the nucleon trajectory, that correspond to lighter states and
thus shall comprise most of the experimental points. Explicitly, we
assign the $N$-resonance states to the Regge trajectories with
different $N$, $n$ and parity as follows
$$
\begin{array}{llll}
  n=0, & N=0, & P=(-1)^{J+N-1/2}: & N(940), \, N_{3/2-}(1520), \, N_{5/2+}(1680),   \\
       &      &                    &  N_{7/2-}(2190), \, N_{9/2+}(2220); \\
  n=0 & N=1, & P=(-1)^{J+N-1/2}: &  N_{1/2-}(1535), \, N_{3/2+}(1720); \\
  n=0 & N=0, & P=(-1)^{J+N+1/2}: & N_{1/2-}(1650), \, N_{3/2+}(1900), \, N_{5/2-}(2200); \\
  n=0, & N=2, & P=(-1)^{J+N-1/2}: & N_{1/2+}(1710), \, N_{3/2-}(2080);\\
  n=0 & N=3, & P=(-1)^{J+N-1/2}: &  N_{1/2-}(2090);\\
  n=1, & N=0, & P=(-1)^{J+N-1/2}: & N_{1/2+}(2100).\\
\end{array}
$$

The calculated Regge trajectories are presented in Figs.~1 and 2. In
Fig.~1 we plot the trajectories of $1/2^+$, $3/2^-$,... states: the
nucleon Regge trajectory, its successor with $N=2$ and the first
radially excited trajectory with $n=1$, $N=0$. Fig.~2 shows the
opposite-parity states: the nucleon successors with $N=1$ and $N=3$
along with the ground-state trajectory ($n=0$, $N=0$) that was
calculated using the opposite-parity formula (\ref{eq302}). The
experimental masses are taken from Particle Data Group \cite{pdg}.
We display the experimental uncertainties if they are high enough.
The unclear states for which the approximate masses are only known
are depicted by empty circles. From Figs.~1 and 2 one can see that
on the plotted trajectories all the $J=1/2$ and $J=3/2$ states below
$2100$~MeV correspond to certain experimentally observed resonances.
In particular, the $N=2$ and $N=3$ trajectories seem to contain
$N_{3/2-}(2080)$ and $N_{1/2-}(2090)$ respectively. In its turn, the
radially excited trajectory in Fig.~1 passes through
$N_{1/2+}(2100)$.

On the other hand, several states drop out of our systematics. These
are the Roper resonance $N_{1/2+}(1440)$,
 $N_{5/2-}(1675)$ and $N_{9/2-}(2250)$. Actually, the position of the Roper resonance is
a longstanding problem since both the conventional three-quark model
\cite{Isgur} and quark-diquark scheme \cite{anisovichb} treat it as
the radial excitation which is unexpected to be lighter than the
first negative-parity state $N_{1/2-}(1535)$. To reproduce the
correct splitting between $N_{1/2+}(1440)$ and $N_{1/2-}(1535)$, the
effects of quark-antiquark pair contributions \cite{Nunez} and of
curvature in combination with the approximate conformal symmetry
\cite{KirchbachCompean} were invoked. Noticeably, all the three
states that are dropped out have masses close to those of the states
with the same spin and opposite parity, $N_{1/2-}(1535)$,
$N_{5/2+}(1680)$ and $N_{9/2-}(2220)$. Thus, in principle, we were
able to incorporate the absent states in our description if we
assumed that the diquark configuration with opposite parity,
\textit{i.e.} pseudoscalar could occur along with the scalar one at
energies above $1400$~MeV. Then the degenerate opposite-parity
energy levels would emerge: $N_{1/2+}(1440)$ and $N_{1/2-}(1535)$
\textit{etc}. However, the accurate treatment of different diquark
configurations should incorporate mixing effects \cite{Nagata} that
is out of scope of the present work.

A few comments on the parity degeneracy are in order. The systematic
parity doubling in excited baryons is usually thought to be a
manifestation of effective chiral symmetry restoration in the upper
part of the spectrum. However, when treated in different approaches,
this phenomenon leads to different multiplet structures of baryon
states (see reviews \cite{jaffe,DoublAfonin}). Under assumption that
the $N$ and $\Delta$-resonances fill out the irreducible
representations of the parity-chiral group $SU(2)_L\otimes
SU(2)_R\otimes C_i$ \cite{DoublGlozmanCohen1}, there appear doublets
of the $N$-resonance states (along with multiplets containing also
$\Delta$'s). Alternatively, the $O(4)\otimes SU(2)_I$ symmetry
advocated in Ref.\cite{o4} implies that $N$'s fall into the
Rarita-Schwinger-like Lorentz multiplets
$(K/2,K/2)\otimes[(1/2,0)\oplus(0,1/2)]$ whose dimensionality starts
with 3, that is the approximately degenerate triplet
$N_{1/2+}(1440)$, $N_{1/2-}(1535)$ and   $N_{3/2-}(1520)$. Note that
our model relies on the Dirac field transforming according to
$(1/2,0)\oplus(0,1/2)$ rather than the Rarita-Schwinger field and
thus has nothing to do with the above-discussed multiplets.

\begin{table}\label{tab1}
\caption{Comparison between the calculated masses of nucleon
resonances and the experimental masses \cite{pdg}. This work:
Eqs.~(33)-(35), FK: AdS/QCD model by Forkel and Klempt \cite{forkel}
(values are cited from \cite{klempt}), CI: Capstick and Isgur
\cite{CI}, BIL: Bijker, Iachello and Leviatan \cite{bijker}, S:
quark-diquark model by Santopinto \cite{santopinto}, BnA and BnB:
Bonn model \cite{Bn}, MK: Skyrme model by Karliner and Mattis
\cite{MK}. Blanks in "This work" column are the states that cannot
be described in the scalar diquark configuration considered (see
Section \ref{part5} for discussion).}
\begin{tabular}{ c c c c c c c c c c}
\hline
Resonance &  Exp  & This work  & FK & CI & BIL & S & BnA & BnB & MK\\
\hline
$N(940)$ & $940$ & $940$   & $943$ & $960$ & 939 & 940 & $939$ & $939$ & $1190$\\
$N_{1/2+}(1440)$ & $1445\pm25$ & & $1396$ & $1540$ & $1444$ & $1562$ & $1698$ & $1540$ &\\
$N_{1/2-}(1535)$ & $1535\pm10$ & $1424$  & $1516$ & $1460$ & $1563$ & $1538$ & $1435$ & $1470$ & $1478$\\
$N_{1/2-}(1650)$ & $1655\pm15$ & $1705$ & $1628$ & $1535$ & 1683 & 1675 & $1660$ & $1767$ \\
$N_{1/2+}(1710)$ & $1710\pm30$ & $1735$  & $1735$ & $1770$ & $1683$ & $1640$ & $1729$ & $1778$ & $1427$\\
$N_{1/2-}(2090)$ & $\approx 2090$ & $1983$  & $2102$ & $2135$ & & & $2200$ & $2180$\\
$N_{1/2+}(2100)$ & $\approx 2100$  & $2098$ & $2017$ & $1975$ & & & $2127$ & $2177$\\

$N_{3/2-}(1520)$ & $1520\pm5$ & $1424$   & $1516$ & $1495$ & $1563$ & $1538$ & $1476$ & $1485$ & $1715$\\
$N_{3/2+}(1720)$ & $1725\pm25$ & $1735$  & $1735$ & $1795$ & $1737$ & $1675$& $1688$ & $1762$ & $1982$\\
$N_{3/2+}(1900)$ & $\approx 1900$ & $2005$  & $1926$ & $1870$ & & & $1899$ & $1904$\\
$N_{3/2-}(2080)$ & $\approx 2080$ & $1983$   & $2102$ & $2125$ & & & $2079$ & $2095$\\

$N_{5/2-} (1675)$ & $1675\pm5$ & & $1628$ & $1630$ & $1683$ & $1671$ & $1655$ & $1622$ & $1744$\\
$N_{5/2+}(1680)$ & $1685\pm5$ & $1735$  & $1735$ & $1770$ & $1737$ & $1675$ & $1723$ & $1718$ & $1823$\\
$N_{5/2-}(2200)$ & $\approx 2200$  & $2252$ & $2102$ & $2234$ & & & $2185$ & $2217$\\
$N_{7/2-}(2190)$ & $2150\pm50$ & $1983$  & $2102$ & $2090$ & $2140$ & & $2093$ & $2100$ & $2075$\\
$N_{9/2+}(2220)$ & $2250\pm50$ & $2196$  & $2265$ & $2327$ & $2271$ & &$2221$ & $2221$ & $2327$\\
$N_{9/2-} (2250)$ & $\approx 2250$ & & $2184$ & $2234$ & $2229$ & & $2212$ & $2170$ & $2234$\\
\hline
\end{tabular}

\end{table}

In Table 1 we list the low-lying states predicted by the present
model as well as by some other models. When comparing these results,
one should keep in mind that the number of fitting parameters ranges
from two in the present, AdS/QCD \cite{forkel} and Skyrme \cite{MK}
models to seven in the relativized model \cite{CI}.

 Inspecting Table 1, one observes degenerate states with increasing
$J$ among the model predictions. The high degeneracies that occur
within the present approach and the AdS/QCD model \cite{forkel}
deserve some explanation. Within the latter model there is
$(\ell+n)$-degeneracy where $\ell$ and $n$ are the orbital and
radial quantum numbers respectively. This implies that intrinsic
orbital and spin angular momenta can be assigned to the observed
states -- the assumption that is feasible since the spin-orbital
coupling is small for baryons. In contrast, our model conserves the
total angular momentum $J$ and its part $N$, which is different from
the orbital and spin ones and stems from the specific oscillator
interaction. The resulting $(J+N)$-degeneracy is a bit weaker than
the $(\ell+n)$-degeneracy of the AdS/QCD approach. For example, our
model predicts that $N_{5/2-}(2200)$ should be substantially heavier
than $N_{1/2-}(2090)$, $N_{3/2-}(2080)$ and $N_{7/2-}(2190)$ which
are degenerate in both the models.

\section{Conclusion}
\label{part6}

In this work the exactly solvable three-body oscillator model with
the spin-1/2 content has been constructed by employing the extension
of the $SL(2,C)$ group to the $Sp(4,C)$ one. The Dirac-like equation
for the $Sp(4,C)$ spinors incorporates the ordinary relativistic
kinematics, but in the presence of interaction differs significantly
from the equations of the other three-body oscillator models, in
particular, of the FKR model \cite{feynman}. The main feature is
that the present model includes only one three-dimensional
oscillator mode, whereas the remaining degrees of freedom of
relative motion are spent to get the rotational excitations. The
corresponding quantum number $N$ goes as the addition to the total
spin $J$, so that the energy spectrum possesses the 
$(J+N)$-degeneracy. The application to the nucleon resonance mass
spectrum has shown that such a model results in the Regge
trajectories $J(E^2)$ that are asymptotically linear in $E^2$  and
do not contain any missing states with $J=1/2$ and $J=3/2$ states
below $2100$~MeV. 
Several states such as the Roper resonance drop out of the developed
simplistic model, which has only two parameters, the oscillator
coupling constant and the nucleon groud-state mass, and uses the
scalar diquark configuration solely. Some modifications of the
model, in particular, introducing extra interactions may be needed
to reproduce the electric form-factors that will be discussed
elsewhere.

\section*{Acknowledgements}
\label{part7}
We thank the anonymous referees for useful suggestions aimed at improving the paper.

\begin{figure}[htbp]
\centerline{\includegraphics[scale=0.44]{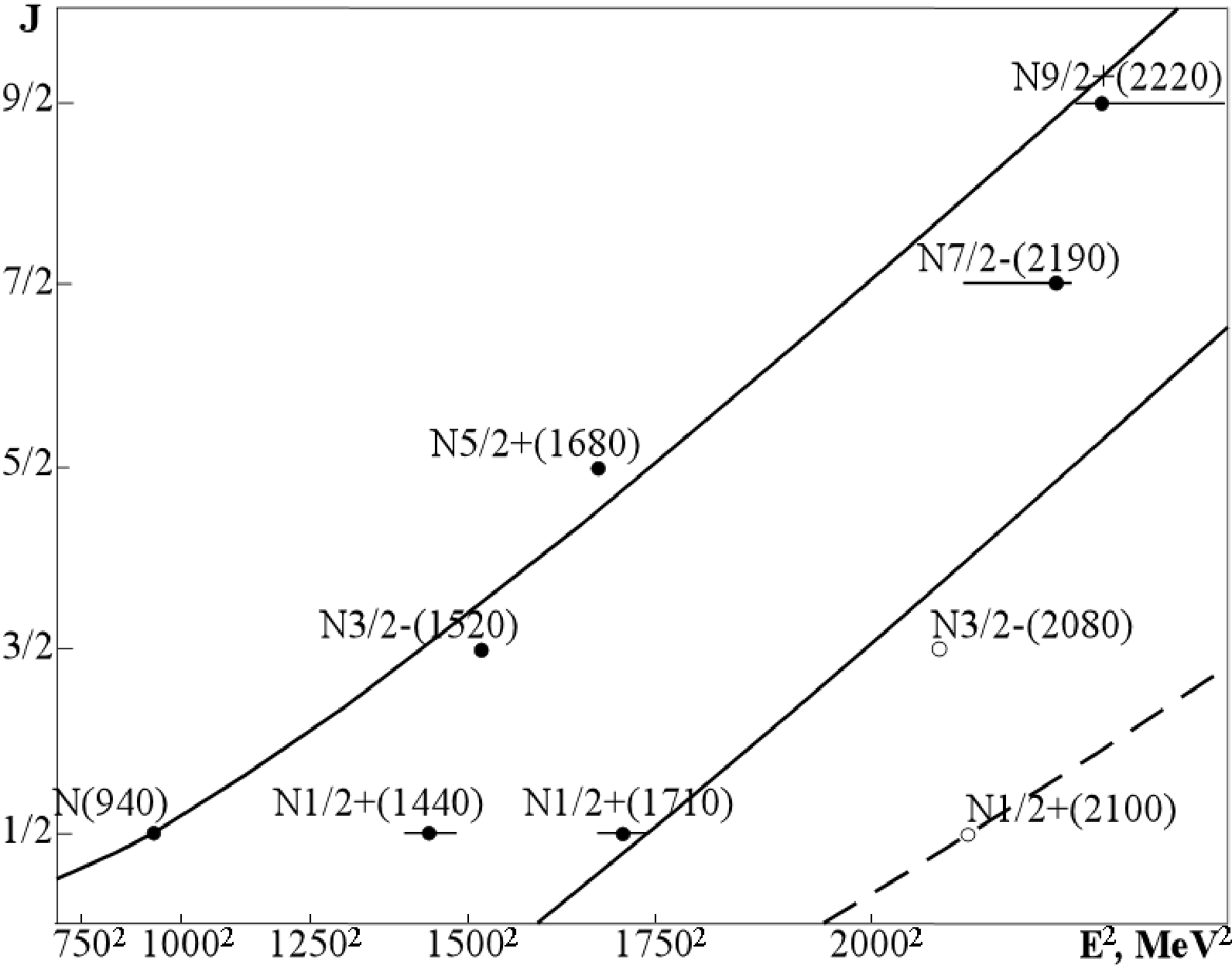}}
\label{fig1} \small  \caption{Nucleon resonance Regge trajectories
that start with $J^P={1/2}^{+}$, calculated using Eqs.~(33) and (34) with
$m=0.940\,$GeV, $\lambda=0.345\,$GeV$^2$. The solid lines are
obtained with the quantum number values $n=0$, $N=0$ and $N=2$,
the dashed line corresponds to $n=1$, $N=0$. The experimental
masses and errors are taken from \cite{pdg}. \hfill} \vspace{20pt}
\centerline{\includegraphics[scale=0.44]{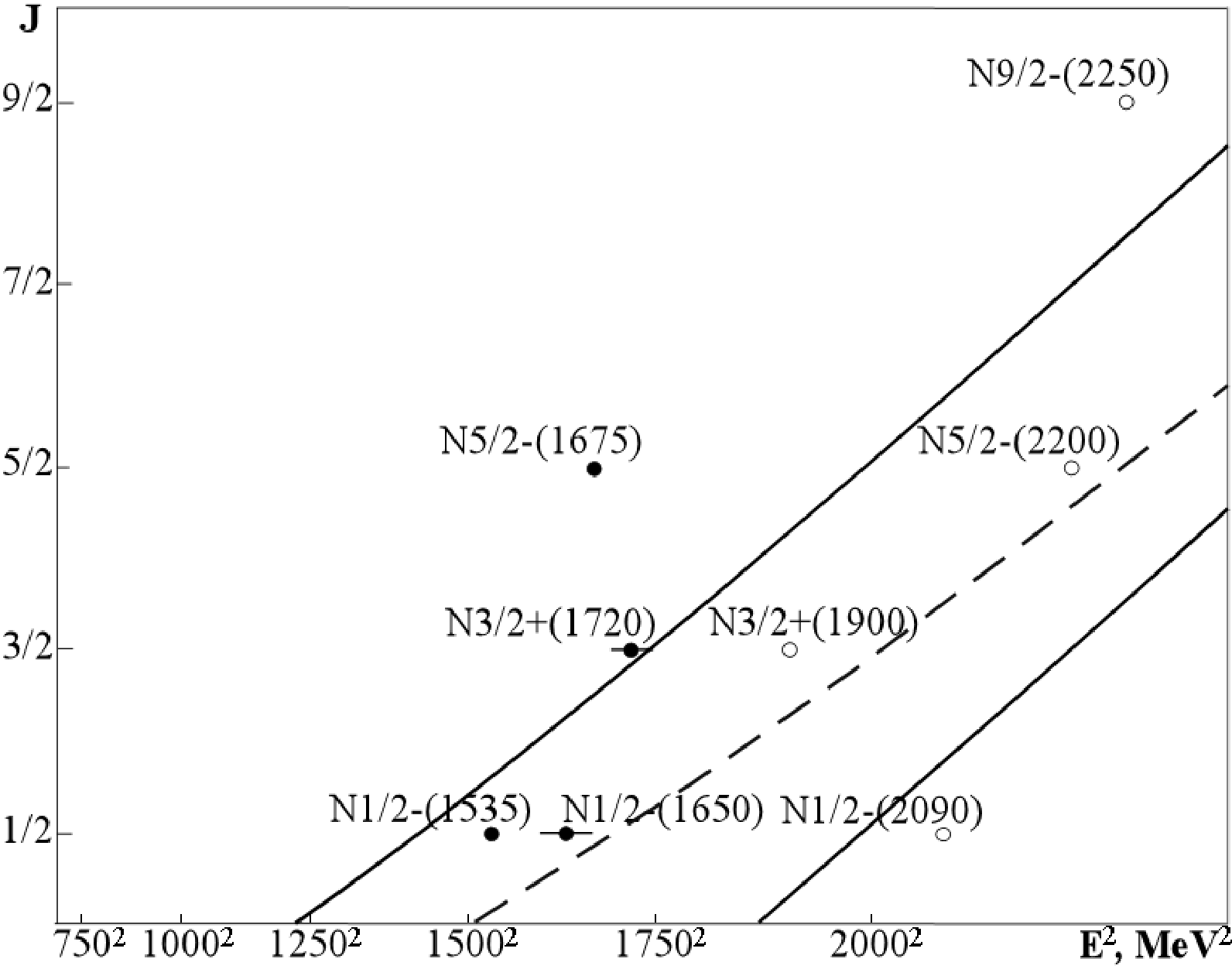}}
\label{fig2} \small  \caption{Nucleon resonance Regge trajectories
that start with $J^P={1/2}^{-}$, calculated using the same
parameter values  as for Fig.~1. The solid lines are obtained from
Eq.~(33) with $n=0$, $N=1$ and $N=3$, the dashed line results
from Eq.~(35) with $n=0$, $N=0$. \hfill}
\end{figure}


\begin{thebibliography}{00}


\bibitem{feynman} R. P. Feynman, M. Kislinger, F. Ravndal, Phys. Rev. D 3, 2706 (1971)

\bibitem{mitra} A. N. Mitra, Phys. Rev. D 11, 3270 (1975)

\bibitem{toki} H. Toki, J. Dey, M. Dey, Phys. Lett. B 133, 20 (1983)

\bibitem{bohm} A. Bohm, M. E. Loewe, P. Magnollay, Phys. Rev. D 32, 791 (1985)

\bibitem{ishida} S. Ishida, K. Yamada, Progr. Theor. Phys. 91, 775 (1994)

\bibitem{semay} F. Buisseret, C. Semay, Phys. Rev. D 73, 114011 (2006)


\bibitem{bijker} R. Bijker, F. Iachello, A. Leviatan, Annals of Phys. 284, 89 (2000)

\bibitem{o4} M. Kirchbach, M. Moshinsky, Yu. F. Smirnov, Phys. Rev. D 64, 114005 (2001)

\bibitem{moshinsky} M. Moshinsky, A. Szczepaniak, J. Phys. A 22, L817 (1989)


\bibitem{kulikov} D. A. Kulikov, R. S. Tutik, A. P. Yaroshenko,
Phys. Lett. B 644, 311 (2007)






\bibitem{landau} E. M. Lifshitz, L. P. Pitaevskii,
V. I. Berestetskii, Quantum Electrodynamics (Pergamon,
Oxford, 1982)


\bibitem{droz06} Ph. Droz-Vincent, Phys. Rev. A
73, 042101 (2006)




\bibitem{elliott} J. P. Elliott, P. G. Dawber, Symmetry in Physics,
Vol.~2 (Oxford University Press, New York, 1986)

\bibitem{Selem} A. Selem, F. Wilczek, arXiv: hep-ph/0602128 (2006)

\bibitem{diquark} M. Anselmino, E. Predazzi, S. Ekelin,
S. Fredriksson, D.B. Lichtenberg, Rev. Mod. Phys. 65, 1199 (1993)

\bibitem{anisovichb} A. V. Anisovich,  V. V. Anisovich,  M. A. Matveev, V. A.
Nikonov, A. V. Sarantsev,  T. O. Vulfs, Int. J. Mod. Phys.
A 25, 2965 (2010)



\bibitem{pdg} J. Beringer et al. (Particle Data Group), Phys. Rev. D 86, 010001 (2012)

\bibitem{Isgur} N. Isgur, G. Karl, Phys. Rev. D
19, 2653 (1979)

\bibitem{Nunez}  M. Nun\~ez V., S. Lerma H., P. O. Hess, S. Jesgarz,
O. Civitarese and M. Reboiro, Phys. Rev. C 70, 025201 (2004)

\bibitem {KirchbachCompean} M. Kirchbach, C. B. Compean,
Phys. Rev. D 82, 034008 (2010)

\bibitem {Nagata} K. Nagata, A. Hosaka,
J. Phys. G 32, 777 (2006)


\bibitem{jaffe} R. L. Jaffe, D. Pirjol, A. Scardicchio,
Phys. Rept. 435, 157 (2006)

\bibitem{DoublAfonin} S. S. Afonin, Int. J. Mod. Phys. A
22, 4537 (2007)

\bibitem{DoublGlozmanCohen1} T. D. Cohen, L. Y. Glozman, Phys. Rev. D 65, 016006 (2002)



\bibitem{forkel} H. Forkel, E. Klempt, Phys. Lett. B 679, 77 (2009)

\bibitem{klempt} E. Klempt, Chinese Phys. C 34, 1241 (2010)

\bibitem{CI} S. Capstick, N. Isgur, Phys. Rev. D 34, 2809 (1986)

\bibitem{santopinto} E. Santopinto, Phys. Rev. C
72, 022201 (2005)

\bibitem{Bn}U. L\"{o}ring, B. Ch. Metsch, H. R. Petry, Eur. Phys. J. A 10, 447 (2001)

\bibitem{MK} M. Karliner, M. P. Mattis, Phys. Rev. D 34, 1991 (1986)




\end{thebibliography}
\end{document}